\documentclass[letterpaper, 10 pt, conference]{ieeeconf}  
\usepackage{amssymb}
 \usepackage[pdftex]{graphicx}

\usepackage{pifont}

\usepackage{multirow}
\usepackage[subrefformat=parens]{subcaption}
\usepackage[utf8]{inputenc}
\usepackage[english]{babel}
 
\usepackage[usenames, dvipsnames]{color}

\usepackage{amsmath}
\usepackage[linesnumbered,ruled]{algorithm2e}

\IEEEoverridecommandlockouts                              

\begin{document}







%

\title{\LARGE \bf
Co-simulation Platform for Developing\\InfoRich Energy-Efficient Connected and Automated Vehicles
}

\author{Shunsuke Aoki$^{1}$, Lung En Jan$^{1}$, Junfeng Zhao$^{2}$, Anand Bhat$^{1}$,\\
Ragunathan (Raj) Rajkumar$^{1}$, Chen-Fang Chang$^{2}$
\thanks{$^{1}$Shunsuke Aoki, Lung En Jan, Anand Bhat, and Ragunathan (Raj) Rajkumar are with Department of Electrical \& Computer Engineering, Carnegie Mellon University
        {\tt\small \{shunsuka, ljan, anandbha, rajkumar\}@andrew.cmu.edu}}%
\thanks{$^{2}$Junfeng Zhao and Chen-Fang Chang are with General Motors Company
        {\tt\small \{junfeng.zhao, chen-fang.chang\}@gm.com}}%
}


\maketitle
\thispagestyle{empty}
\pagestyle{empty}

\begin{abstract}

With advances in sensing, computing, and communication technologies, Connected and Automated Vehicles (CAVs) are becoming feasible. The advent of CAVs presents new opportunities to improve the energy efficiency of individual vehicles.
However, testing and verifying energy-efficient autonomous driving systems are difficult due to safety considerations and repeatability.
In this paper, we present a co-simulation platform to develop and test novel vehicle eco-autonomous driving technologies named {\it InfoRich}, which incorporates the information from on-board sensors, V2X communications, and map database.
The co-simulation platform includes eco-autonomous driving software, vehicle dynamics and powertrain (VD\&PT) model, and a traffic environment simulator.
Also, we utilize synthetic drive cycles derived from real-world driving data to test the strategies under realistic driving scenarios.
To build road networks from the real-world driving data, we develop an Automated Parser and Calculator for Map/Scenario named {\it AutoPASCAL}.
Overall, the simulation platform provides a realistic vehicle model, powertrain model, sensor model, traffic model, and road-network model to enable the evaluation of the energy efficiency of eco-autonomous driving.

\end{abstract}

\section{INTRODUCTION}


Energy efficiency is one of the most significant factors for vehicles.
According to the U.S. Energy Information Administration (EIA), the U.S. net import of petroleum is approximately equal to 11 \% of U.S. petroleum consumption. In addition, $CO_2$ emissions due to motor gasoline and diesel fuel consumption is over 30 \% of total U.S. energy-related $CO_2$ emissions in 2017 \cite{US_EIA}.

The advent of Connected and Automated Vehicles (CAVs) is an opportunity to improve not only safety and human comfort \cite{hemmati2019adaptive, aoki2018dynamic}, but also vehicle fuel economy.
We work on designing and testing a novel vehicle dynamics and powertrain (VD\&PT) technology named {\it InfoRich}, which aims to improve fuel efficiency by incorporating the on-board sensors, V2X (Vehicle-to-Everything) communications, and map database.


One of the significant challenges of developing CAV technologies is testing, verifying, and validating the eco-driving technologies in a real-world traffic environment because of cost, safety, and repeatability.
First, testing self-driving vehicles in the real world is costly in terms of time and resources. 
Secondly, misbehavior on public roads can lead to safety issues.
Thirdly, since traffic situations change in the real world, a comparative study for multiple schemes is difficult.

\begin{figure}[!b]
  \begin{center}
    \begin{tabular}{c}
      \begin{minipage}{0.45\hsize}
        \begin{center}
          \includegraphics[width=4.05cm]{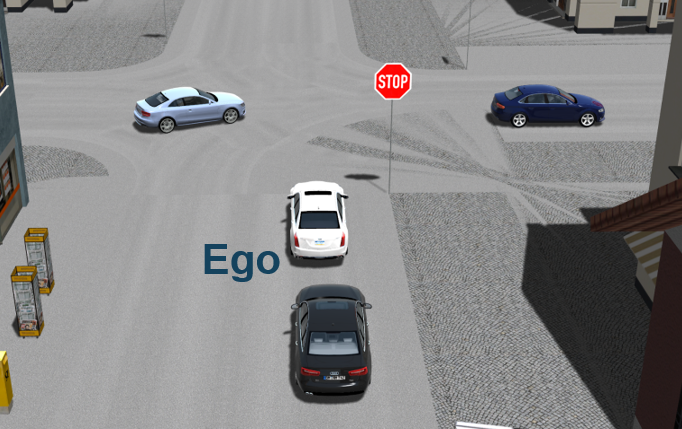}
	\hspace{7.2cm} (a) Moving Obstacles \\ in Traffic Simulator.
        \label{fig:Synchro_2size}
        \end{center}
      \end{minipage}
      \begin{minipage}{0.55\hsize}
        \begin{center}
          \includegraphics[width=4.05cm]{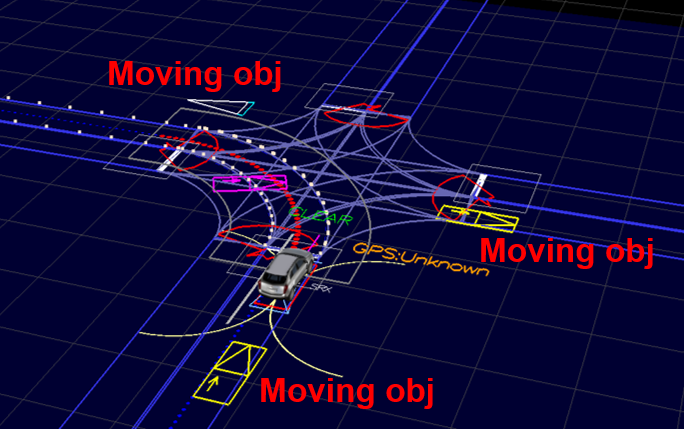}
	\hspace{7.2cm} (b) Moving Obstacles \\ in iREAD System.
          \label{fig:iread_moving}
        \end{center}
      \end{minipage}\\ \\

      \begin{minipage}{0.45\hsize}
        \begin{center}
          \includegraphics[width=4.05cm]{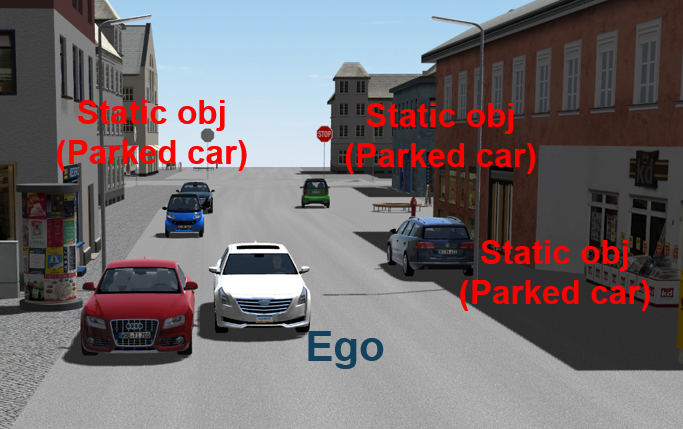}
	\hspace{6.8cm} (c) Static Obstacles \\ in Traffic Simulator.
        \label{fig:vires_static}
        \end{center}
      \end{minipage}
      \begin{minipage}{0.54\hsize}
        \begin{center}
          \includegraphics[width=4.05cm]{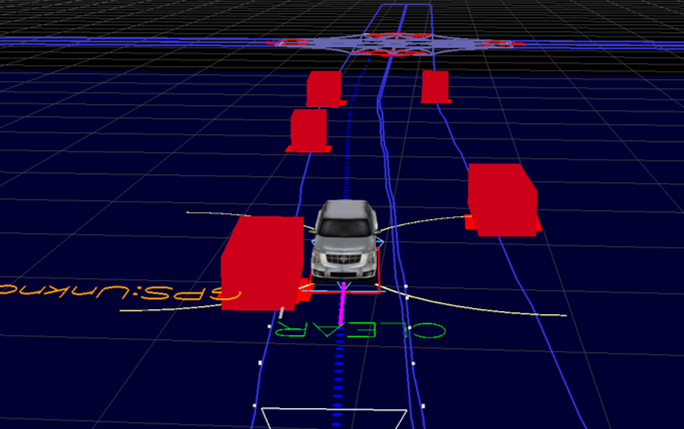}
	\hspace{6.8cm} (d) Static Obstacles \\ in iREAD System.
          \label{fig:iread_static}
        \end{center}
      \end{minipage}
    \end{tabular}
    \caption{InfoRich Co-simulation Environment.}
\label{fig:overview_co-simsystems}
  \end{center}
\end{figure}

In this paper, we present a co-simulation platform architecture to develop, test, and verify eco-autonomous driving technologies, as shown in Figure \ref{fig:overview_co-simsystems}.
The co-simulation platform contains three components: the InfoRich Eco-Autonomous Driving (iREAD) System, a Simulink-based VD\&PT model, and a traffic environment simulator called VIRES Virtual Test Drive (VIRES VTD) \cite{viressimulator, zhao2019virtual}. 
The iREAD system is a hybrid emulator and simulator for autonomous driving vehicles, and it supports eco-driving applications, including Eco-Approach, Eco-Departure, and Eco-Cruise.
The Simulink-based VD\&PT model enables us to test our strategies with a realistic vehicle dynamics and powertrain model.
The traffic environment simulator provides virtual sensors and traffic environments, including traffic models, pedestrian models, static objects, and traffic lights.
The system enables us to test control and planning strategies under a variety of traffic conditions safely, and also enables us to assess the eco-autonomous driving system in a repeatable manner.

\begin{figure*}[!t]
\centering
\includegraphics[width=12.50cm]{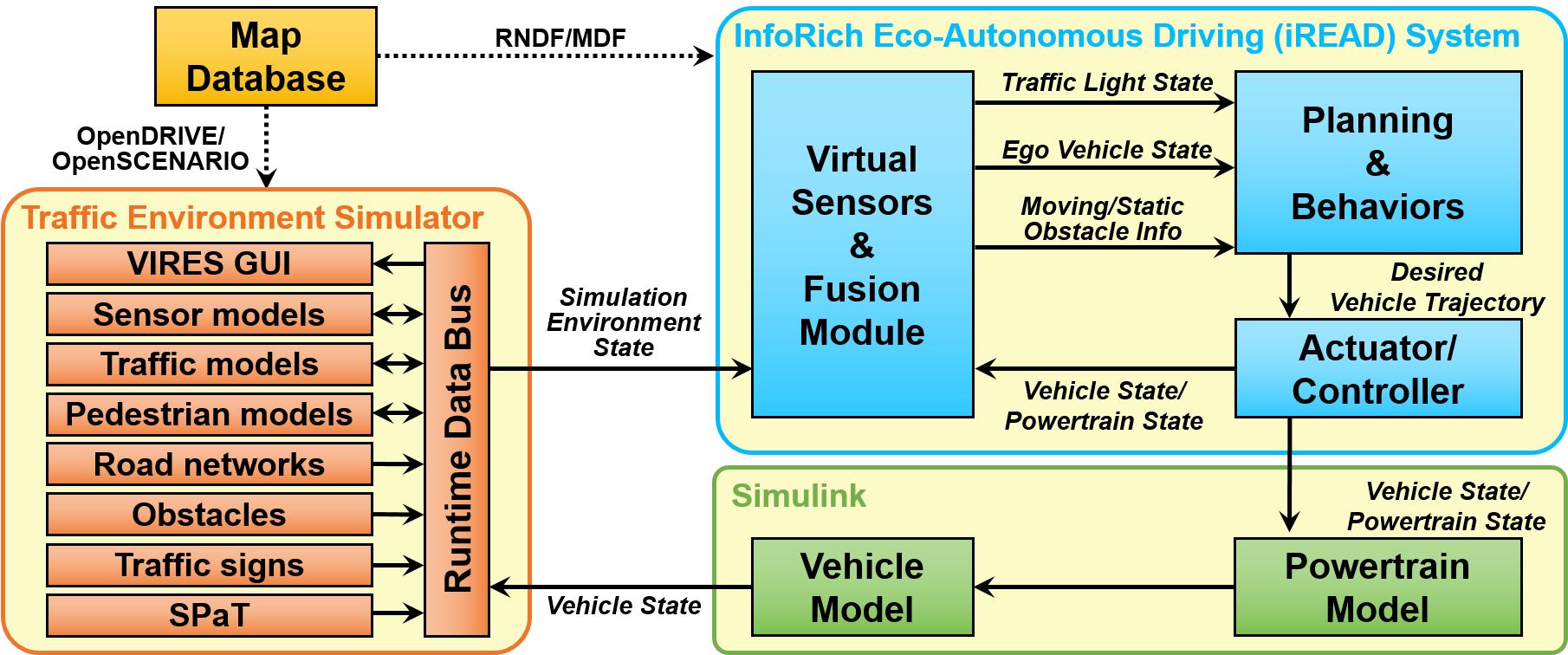}
\caption{Co-simulation Architecture.}
\label{fig:CoSimPlatform}
\end{figure*}

In addition, to simulate and validate the eco-driving control design, we utilize several synthetic drive cycles derived from the Transportation Secure Data Center (TSDC) driving cycle data, which is provided from the National Renewable Energy Laboratory (NREL).
The TSDC data cover over one million miles of driving cycles from dozens of U.S. cities, and this data helps to assess the real-world fuel economy.
The synthetic cycles are pre-processed to remove ties to driver identity information to address privacy concerns.
We also design and develop an Automated Parser and Calculator for Map/Scenario named {\it AutoPASCAL}.
The AutoPASCAL provides map and scenario files for the iREAD system and for the traffic environment simulator while keeping the consistency.
Overall, the co-simulation platform enables us to test and verify the InfoRich eco-autonomous driving system with a realistic vehicle dynamics model, a powertrain model, a traffic model, and a road-network model.


\vspace{0.1in}

The contributions of this paper are as follows:

\begin{enumerate}
 \item We present a co-simulation environment to test, verify, and validate energy-efficient control strategies for connected and automated vehicles.
 \item We demonstrate the utility of synthetic scenarios and {\it AutoPASCAL} to test control and planning strategies under the realistic driving scenarios.
 \item We introduce the {\it InfoRich} driver concept that uses the information from on-board sensors, V2X communications, and map database to reduce the fuel consumption. 
\end{enumerate}

\vspace{0.1in}

The remainder of this paper is organized as follows.
Section I\hspace{-.1em}I discusses previous work related to our research.
Section I\hspace{-.1em}I\hspace{-.1em}I describes the overall co-simulation platform for eco-driving CAVs.
Section I\hspace{-.1em}V presents the real-world driving scenarios we generate, test, and evaluate.
Section V gives the evaluation of baseline driver and the discussion of the eco-driving strategies.
Finally, Section V\hspace{-.1em}I presents our summary and future work.

\vspace{0.1in}

\section{RELATED WORK}

A CAV comprises a large number of hardware and software components, and a well-defined methodology supported by a suite of tools is needed for validation and verification.
To design, develop, test, and validate full self-driving capabilities, Carnegie Mellon University outfitted a Cadillac SRX \cite{wei2013towards} to drive itself and developed a host of tools to aid the testing of autonomous driving systems \cite{bhat2018tools}.
In addition, Autoware project \cite{kato2018autoware} provides an open-source software for autonomous driving systems.
These projects do not provide any tools to simulate a realistic vehicle control model, a powertrain model, or traffic model, that are essential for designing and testing eco-driving strategies and fully supported by the InfoRich co-simulation platform.


To study and understand CAVs, traffic simulators and vehicular ad hoc networks (VANET) simulators are also necessary.
Traffic simulators, such as SUMO \cite{behrisch2011sumo, lopez2018microscopic}, VISSIM \cite{cecchini2017ltev2vsim}, Groovenet \cite{mangharam2006groovenet}, AutoSim \cite{aoki2017merging} and MoVES \cite{bononi2006parallel} simulate realistic vehicular mobility and the connectivity of vehicular communications.
VANET simulators such as TraNS \cite{piorkowski2008trans} and Veins \cite{sommer2011bidirectionally} integrate a mobility generator and a network simulator to provide a tool to study V2X communications.
CARLA \cite{dosovitskiy2017carla} is an open-source simulator for automated vehicles, and it supports the flexible specification of sensor suites and environmental conditions.
These tools simulate only the vehicle's on-board sensors and/or V2X communications, and do not support with a realistic vehicle control model or a powertrain model, which are important to study and understand the energy efficiency of Connected and Automated Vehicles.

Co-simulation systems have been developed for studying system safety in the real world \cite{serban2017co, jung2007modeling, palensky2013modeling}.
R. Serban et al. designed and studied a co-simulation environment for off-road vehicles to understand the interaction between vehicle tires and road surface \cite{serban2017co}.
D. Jung et al. developed a Hardware-in-the-Loop (HIL) simulation environment for Unmanned Aerial Vehicles (UAVs) \cite{jung2007modeling}.
Also, P. Palensky et al. have worked on intelligent energy systems including electric vehicles, and developed a co-simulation environment with multiple components \cite{palensky2013modeling}.
There are multiple studies for co-simulation architectures for vehicles, but to the best of our knowledge, there are no existing systems specific for Connected and Automated Vehicles.

\begin{figure*}[!t]
\centering
\includegraphics[width=17.50cm]{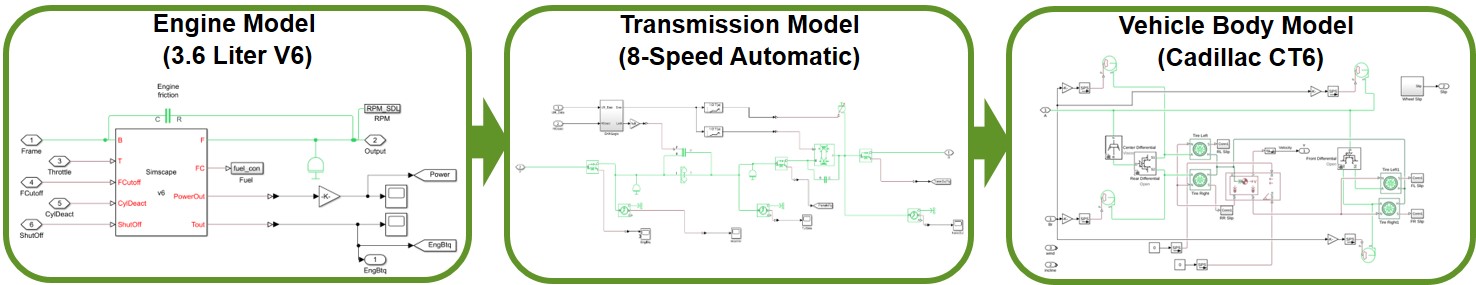}
\caption{VD/PT model for Cadillac CT6.}
\label{fig:powertrain}
\end{figure*}

\vspace{0.1in}

\section{CO-SIMULATION ARCHITECTURE}

This section presents the co-simulation architecture.
As shown in Figure \ref{fig:CoSimPlatform}, the platform is composed of three components: a Simulink-based VD\&PT model, a traffic environment simulator called VIRES Virtual Test Drive (VIRES VTD) \cite{viressimulator}, and the InfoRich Eco-Autonomous Driving (iREAD) System.

In the loop of the co-simulation, the traffic environment simulator transmits the Environment State to the iREAD system, including the Ego Vehicle State, the Traffic Light State, and the Moving/Static Obstacle Information.
The iREAD system calculates and transmits the Vehicle/Powertrain State (e.g. speed, location, and acceleration) and Control information to the Simulink-based model.
The Simulink model provides the Vehicle State to the VIRES VTD traffic environment simulator.

Communications among these three components use UDP (User Datagram Protocol) to increase the system flexibility and the data transmission rate.
The communication frequency for three components is set to 50 Hz, except for the Traffic Light Information from the traffic environment simulator to the iREAD system.
The Traffic Light Information is set to 10 Hz based on the DSRC (Dedicated-Short Range Communications) standard \cite{kenney2011dedicated}.

\begin{figure}[b]
\centering
\includegraphics[width=7.50cm]{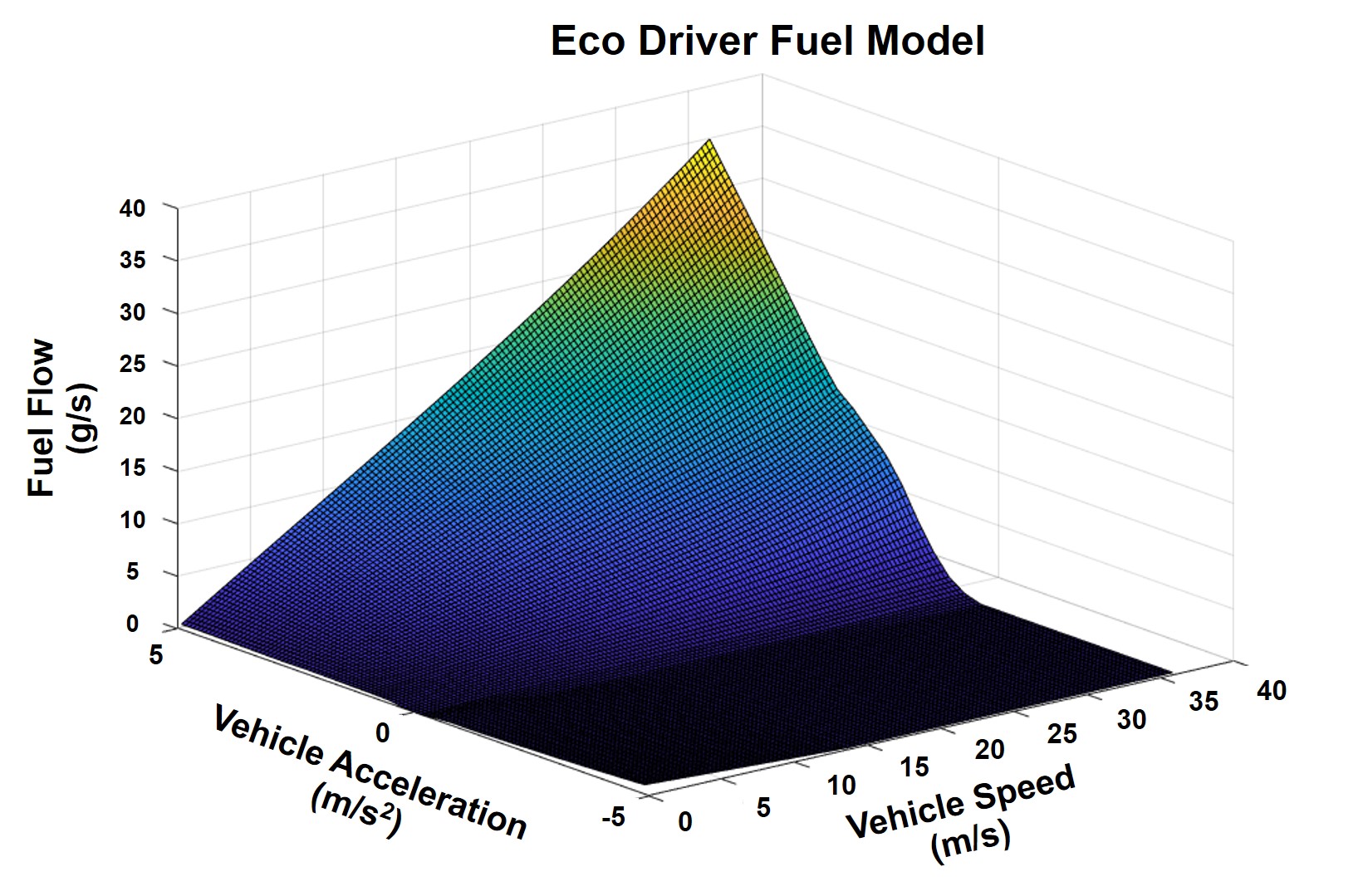}
\caption{Fuel Table for Cadillac CT6.}
\label{fig:fueltable}
\end{figure}

 \subsection{Vehicle Dynamics/Powertrain Model}




In this study, we develop a high-fidelity model for vehicle dynamics and powertrain (VD\&PT) model by using Simulink Simscape Driveline platform.
The Simulink Simscape Driveline platform enables us to construct a network of physical components to model the vehicle systems within the Simulink environment.
The Simulink Simscape has two advantages: (i) enabling to analyze components of the powertrain on an individual basis and (ii) enabling to create custom components through the Simscape language and cross-domain simulation with electrical and thermal systems.
In this paper, we use the vehicle model and data of General Motors Cadillac CT6.


As shown in Figure \ref{fig:powertrain}, the VD/PT model consists of a 3.6-liter V6 engine, torque converter, 8-speed automatic transmission, and vehicle chassis.
We build a custom powertrain model using the Simscape language, whose input signals include throttle position, Deceleration Fuel Cut Off (DFCO), engine shut off, and cylinder deactivation. Engine parameters include torque-speed tables, fuel consumption tables, engine inertia, and idle reference speed.
As one of the exemplifying parameters, Figure \ref{fig:fueltable} presents the fuel-rate lookup tables for the Cadillac CT6.

In addition, we develop the 8-speed transmission model including the gear ratio and inertia based on Cadillac CT6.
The simplified transmission model contains a variable inertia, which assumes different values for each gear.
The torque converter takes speed ratio vector, torque ratio vector, and capacity factor (K factor) vector as its parameters. The transmission upshift and downshift tables as well as the torque converter lockup have also been implemented.

The Simscape model of the vehicle body consists of the vehicle's longitudinal dynamics, tire model, brakes, and all-wheel drive system with the transfer case, as well as front and rear differentials. The vehicle body parameters include frontal area, drag coefficient, vehicle mass, wheel/tire radius, wheel inertia and final drive gear.
The gain for the braking force is tuned using the transmission output torque from the GM data to match the speed profile for the FTP-75 test cycle used by the United States Environmental Protection Agency (EPA) for emissions testing.

 \subsection{VIRES VTD: Traffic Environment Simulator}

VIRES VTD is a tool-chain for the creation, configuration, presentation, and evaluation of virtual traffic environments in the scope of road and rail based simulations.
In the co-simulation platform, as shown in Figure \ref{fig:CoSimPlatform}, the VIRES VTD traffic environment simulator provides 8 features: (i) 3D GUI, (ii) Sensor models, (iii) Traffic models, (iv) Pedestrian models, (v) Road networks, (vi) Static objects, (vii) Traffic signs, and (viii) Signal Phases and Timing (SPaT).


The traffic environment simulator relies on three file formats for traffic environments: OpenDRIVE, OpenSCENARIO, and OpenCRG \cite{dupuis2010opendrive}.
OpenDRIVE represents the logical description of road networks that include road curvatures and grades, lane numbers, road intersections, traffic signs, and speed limits.
OpenSCENARIO is for the description of dynamic contents in simulation, such as surrounding traffic, static obstacles, and Signal Phases and Timing (SPaT).
OpenCRG presents the detailed description of road surfaces, which enables tire-, vibration-, and/or driving-simulations.

In the traffic environment simulator, we configure the coverage ranges and physical locations of the virtual sensors and vehicular communications.
Also, we can change the parameters for traffic environments, such as a traffic density, a pedestrian density, traffic signs, and SPaT for traffic lights.
By changing these parameters, we can test the control and planning strategies under a variety of traffic conditions in a repeatable manner.

For data transmissions and exchanges, the VIRES VTD traffic environment simulator use a Runtime Data Bus. Using the Runtime Data Bus, the run-time data are exchanged between the traffic environment simulator core components and third party tools.

 \subsection{iREAD: Eco-Autonomous Driving System}

iREAD (InfoRich Eco-Autonomous Driving) System is the next generation of TROCS (Tartan Racing On-board Computer System) that is a system-level hybrid emulator and simulator for autonomous vehicles \cite{bhat2018tools, urmson2008autonomous}.
Compared to TROCS, the iREAD system has two key additional features: Virtual Sensors \& Fusion Module and Planning \& Behaviors for eco-driving.


First, in the Sensors \& Fusion Module, the iREAD system parses the sensor data from the traffic environment simulator as if real data from the physical sensors.
Specifically, the iREAD system holds the physical size, speed, acceleration, position, and heading information for each moving or static obstacle in the sensor coverage area.
As shown in Figure \ref{fig:overview_co-simsystems}, iREAD GUI shows the moving and static obstacles that are generated by the VIRES world simulator.


Also, the co-simulation platform supports Vehicle-to-Infrastructure (V2I) Communications.
The vehicular communication range is set to be three hundred meters and the communication frequency is set to 10 Hz  based on the DSRC standard \cite{kenney2011dedicated}.
Each traffic light has the wireless interface and the Ego vehicle can get the SPaT information from the dedicated traffic light(s) within the communication range.

Second, in the Planning \& Behaviors, the iREAD system supports multiple eco-driving applications, including Eco-Approach, Eco-Departure, and Eco-Cruise.
The Eco-Approach is defined as the fuel-efficient vehicle/powertrain operation that aggressively applies coasting to bring a vehicle to stop.
By predicting the stop in advance and proactively engaging in coasting to efficiently convert the vehicle kinematic energy to the distance traveled, we can save the fuel consumption.
Also, the Eco-Departure maneuver should conduct a smooth and less-aggressive acceleration profile to reach the target cruise speed.
The Eco-Cruise might adjust its speed based on traffic, speed limit changes, and navigational maneuvers.

For map and scenario files, the iREAD system relies on extended version of the RNDF (Route-Network Definition Format) and MDF (Mission Definition Format) \cite{challenge2007route}, respectively.
Most of the logical attributes of road networks are stored into RNDF, and the speed limits information are described in MDF.
To keep the consistency between OpenDRIVE/OpenSCENARIO and RNDF/MDF in the co-simulation platform, we developed the Automated Parser and Calculator for Map/Scenario (AutoPASCAL) to be discussed in Section \ref{Map_and_Scenario}.


\vspace{0.1in}

\section{REAL-WORLD DRIVING SCENARIOS}

To simulate and validate our eco-autonomous driving strategies, we utilize several synthetic drive cycles derived from the Transportation Secure Data Center (TSDC) driving cycle data.
In this section, we present the synthetic drive cycles and our AutoPASCAL (Automated Parser and Calculator for Map/Scenario) for the co-simulation platform.

\begin{figure}[!t]
\centering
\includegraphics[width=8.50cm]{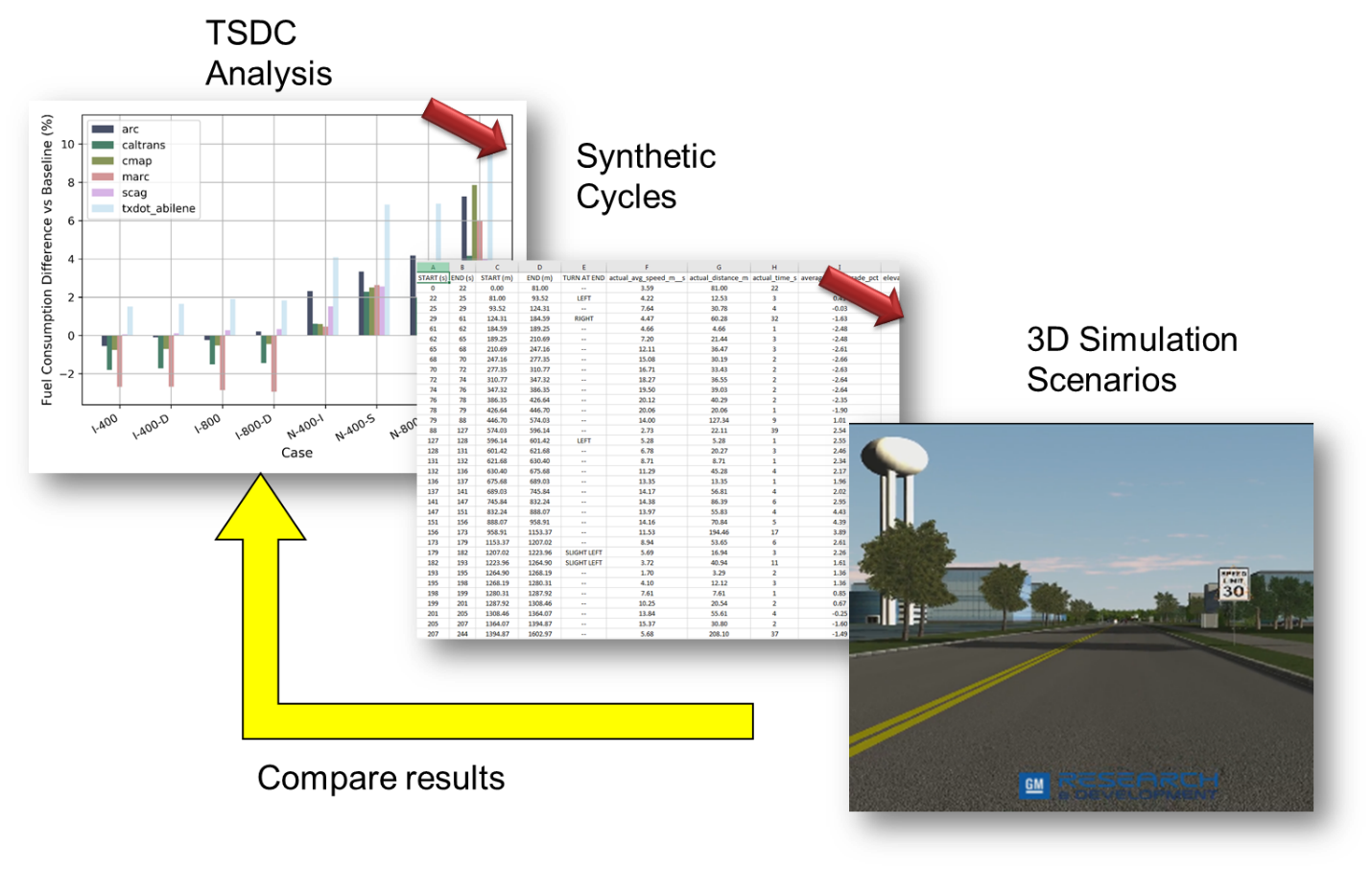}
\caption{Work Flow for Synthetic Scenarios Creation.}
\label{fig:syntheticscenario}
\end{figure}

\begin{figure}[!t]
\centering
\includegraphics[width=8.50cm]{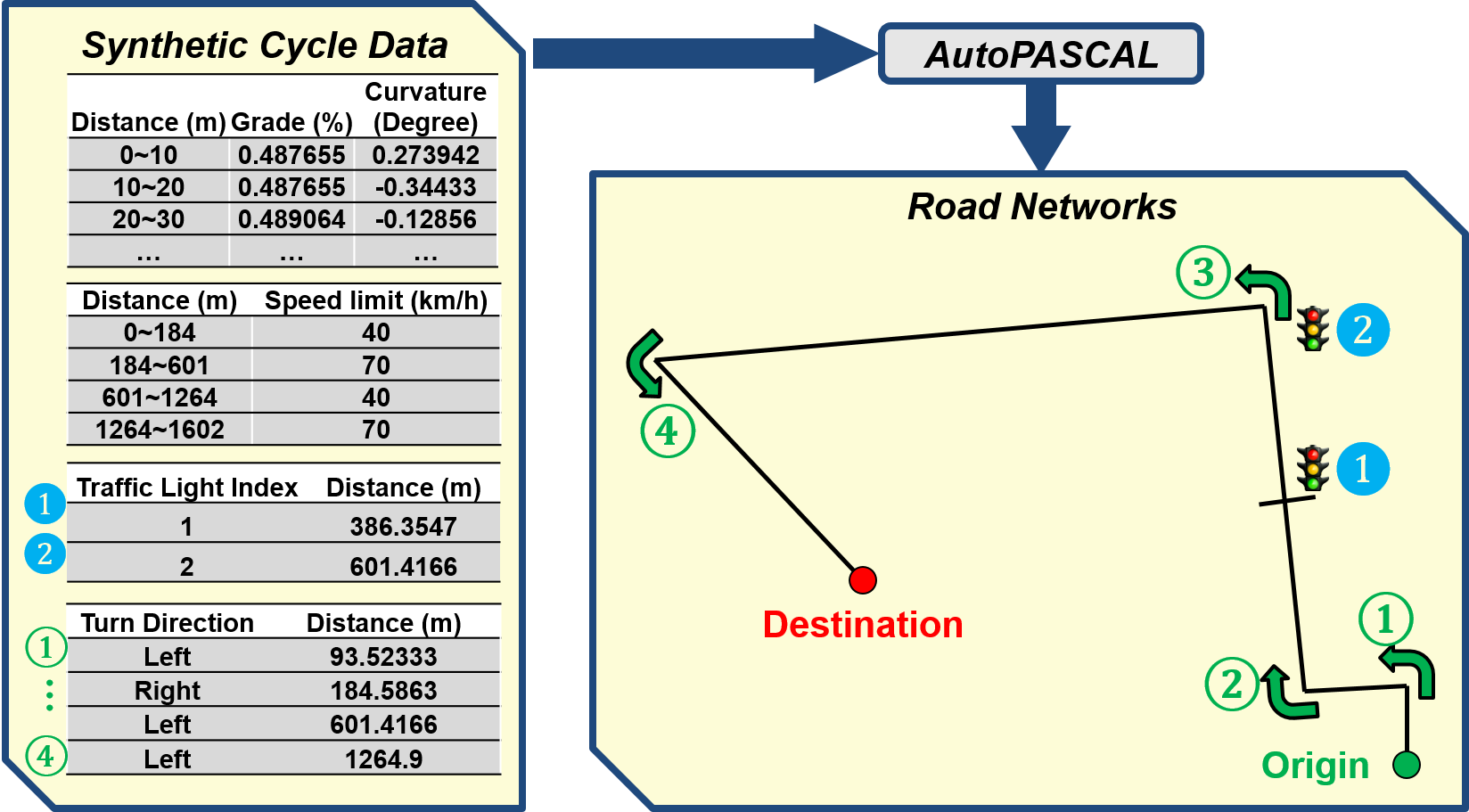}
\caption{Example of Synthetic Cycle with AutoPASCAL.}
\label{fig:Synthetic_TSDC}
\end{figure}

 \subsection{Synthetic Scenarios}

With the synthetic drive data, we can conduct multiple simulation runs in a deterministic and repetitive fashion.
The impact of test factors, such as congestion level and penetration rates, can also be simulated and studied in a more systematic way.
The workflow of creating driving scenarios is shown in Figure \ref{fig:syntheticscenario}:



\begin{enumerate}
 \item Simulated the 79,682 trips in the TSDC.
 \item Tallied fuel savings for Eco-Approach, Eco-Departure, and Eco-Cruise.  
 \item Ranked each trip by total energy savings.  
 \item Identified trips around the 50th, 75th, and 95th percentiles.
 \item Selected 3 “short” trips (approximately 1 mile in length) from each percentile set and 3 “long” trips (approximately 10 miles in length) plus one additional short and one additional long trip to make a total of 20 cycles.
\end{enumerate}


Overall, these trips are able to represent real-world driving with various degrees of predicted energy savings.
The synthetic scenarios have a total length of 176.4 (km), approximately 120 traffic lights, and 280 road intersections.

The synthetic drive data contains a variety of information for each road segment, such as the distance, speed limit, grade, curvature, and traffic controllers.
The synthetic drive data are pre-processed to address privacy concerns, and our AutoPASCAL develops the map and scenario files for the InfoRich co-simulation platform.
In-depth simulation can be conducted on the platform with controlled and detailed 3D virtual environment.
The eco-driving algorithm's performance will be evaluated through the in-depth simulation with much higher resolution.
Then the in-depth simulation results will be compared to the predicted energy savings mentioned above to make sure that the fuel benefits from the validated eco-driving algorithm can be expected in real-world driving.

 \subsection{AutoPASCAL: Automated Parser and Calculator for Map/Scenario} \label{Map_and_Scenario}

To build road networks from the synthetic cycle data, we develop the C/C++-based Automated Parser and Calculator for Map/Scenario (AutoPASCAL).
The AutoPASCAL extracts the locations of traffic lights and road intersections and the variety of information for each road segment as shown in Figure \ref{fig:Synthetic_TSDC}.
Here, the Distance in the synthetic cycle data represents the trip distance from the Origin along the roads to be passed.

\begin{figure}[!t]
\centering
\includegraphics[width=7.70cm]{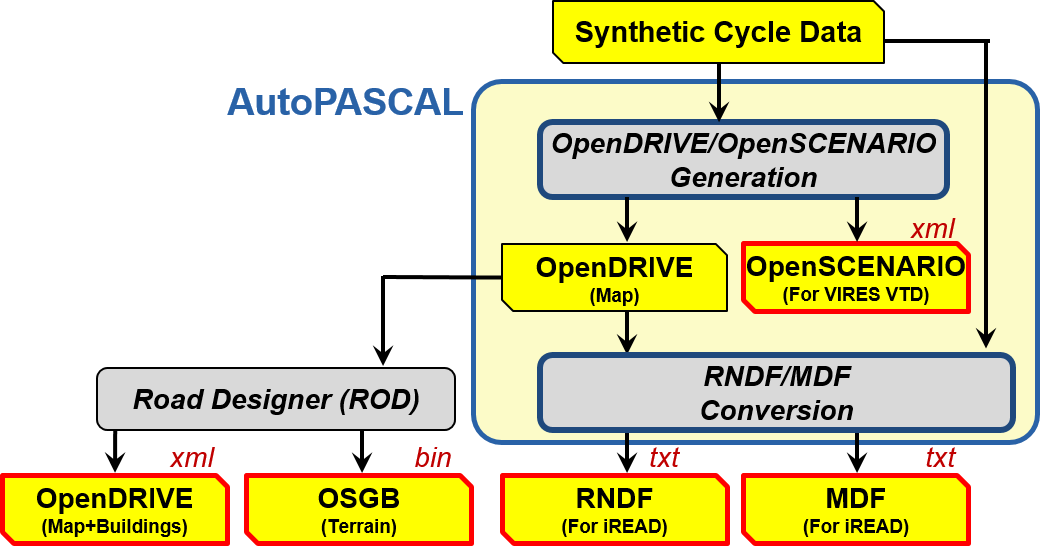}
\caption{File Generation by AutoPASCAL and ROD.}
\label{fig:MapAutomation}
\end{figure}

The AutoPASCAL generates two subsets of map/scenarios files for the co-simulation platform: OpenDRIVE/OpenSCENARIO and RNDF/MDF.
The traffic environment simulator uses OpenDRIVE/OpenSCENARIO and the iREAD system builds the road networks with RNDF/MDF.
The AutoPASCAL keeps the consistency of the geographical features between these files by 2-step data processing as shown in Figure. \ref{fig:MapAutomation}: (i) OpenDRIVE/OpenSCENARIO Generation and (ii) RNDF/MDF Conversion.



\begin{figure}[!b]
  \begin{center}
    \begin{tabular}{c}
      \begin{minipage}{0.5\hsize}
        \begin{center}
          \includegraphics[width=4.0cm]{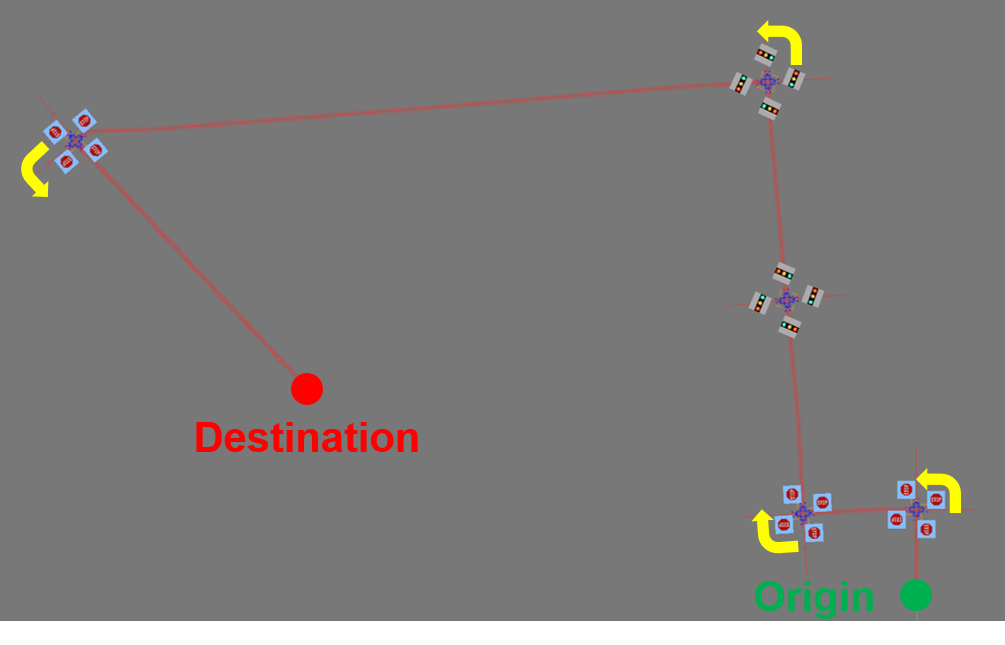}
	\hspace{6.8cm} (a) Visualized\\ in Traffic Simulator.
        \label{fig:vires_static}
        \end{center}
      \end{minipage}
      \begin{minipage}{0.5\hsize}
        \begin{center}
          \includegraphics[width=4.0cm]{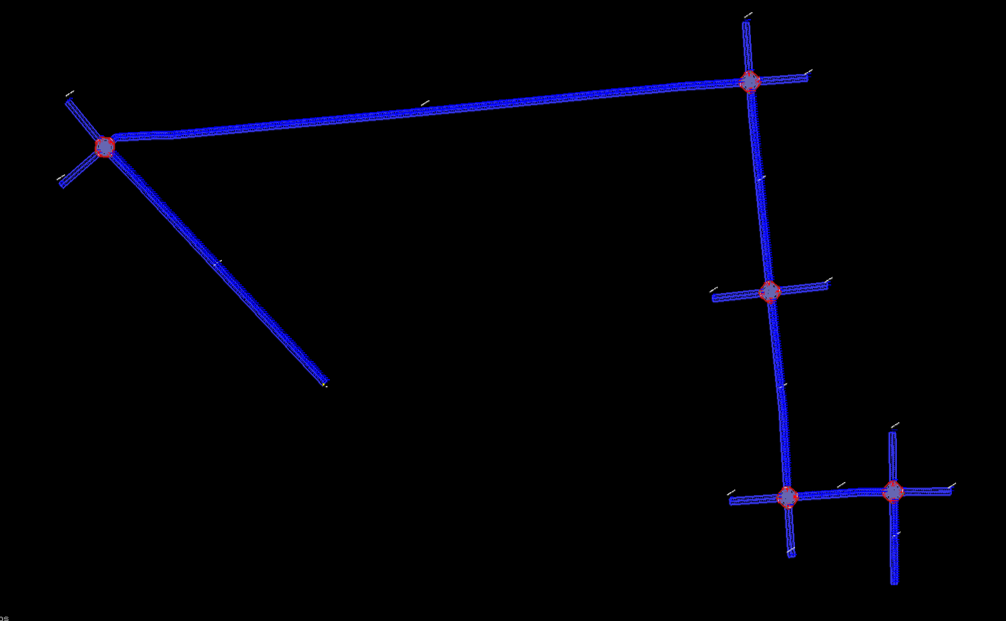}
	\hspace{6.8cm} (b) Visualized\\ in iREAD System.
          \label{fig:iread_static}
        \end{center}
      \end{minipage}
    \end{tabular}
    \caption{A Short Synthetic Scenario.}
\label{fig:NRELA_visualization}
  \end{center}
\end{figure}

\begin{figure}[!b]
  \begin{center}
    \begin{tabular}{c}
      \begin{minipage}{0.5\hsize}
        \begin{center}
          \includegraphics[width=4.0cm]{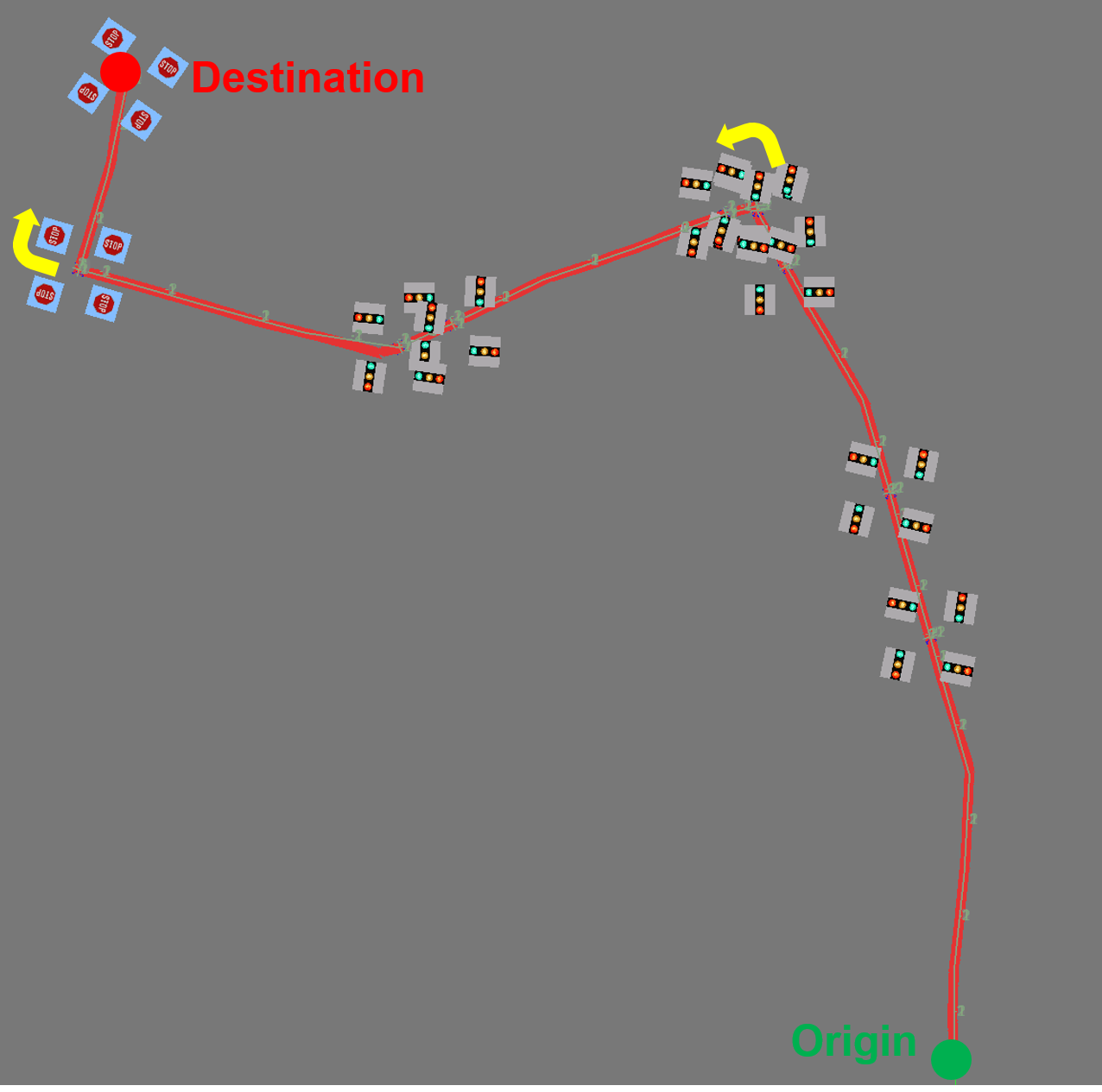}
	\hspace{6.8cm} (a) Visualized\\ in Traffic Simulator.
        \label{fig:vires_static}
        \end{center}
      \end{minipage}
      \begin{minipage}{0.5\hsize}
        \begin{center}
          \includegraphics[width=4.0cm]{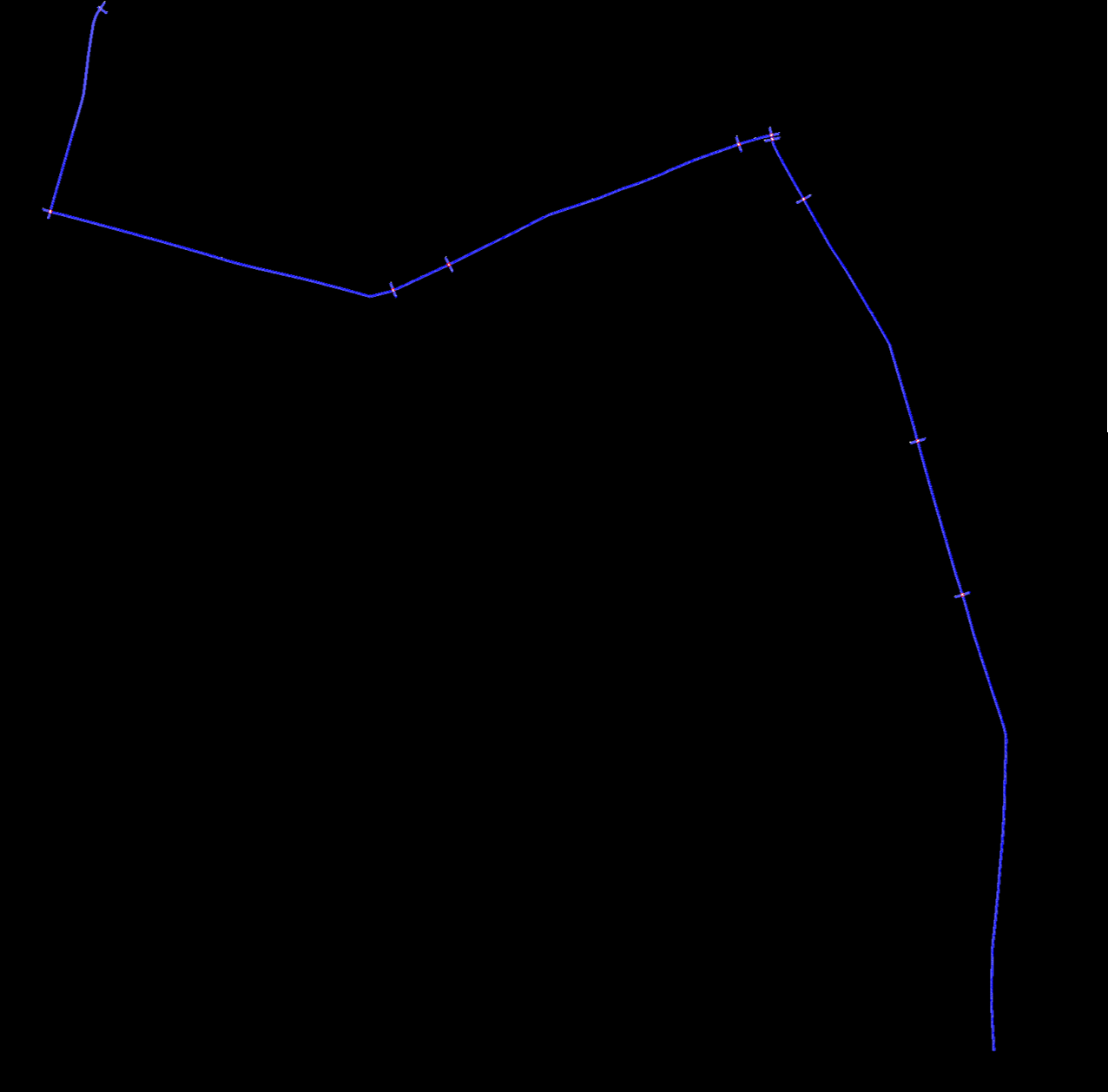}
	\hspace{6.8cm} (b) Visualized\\ in iREAD System.
          \label{fig:iread_static}
        \end{center}
      \end{minipage}
    \end{tabular}
    \caption{A Long Synthetic Scenario.}
\label{fig:NRELB_visualization}
  \end{center}
\end{figure}

\begin{table*}[t]
\centering
\caption{Statistics of Generated Synthetic Cycles.}
\small{
  \begin{tabular}{c|c|c|c|c|c|c|c|c|c} 
                          & Average & Max       & Min        & Average   & Max & Min & Average & Max & Min\\
                          & Distance & Distance & Distance & \# of TLs & \# of TLs& \# of TLs& \# of RIs& \# of RIs& \# of RIs\\ \hline \hline
    10 short cycles & 1.73 (km) & 2.17 (km) & 1.54 (km) & 1.1& 4& 0 & 6& 9& 3\\ \hline 
    10 long cycles & 15.9 (km) & 20.3 (km) & 11.7 (km) & 11& 22& 2 & 21.4& 31& 14\\
  \end{tabular}
}\label{table:NRELB_visualization}
\end{table*}

\begin{enumerate}
  \item {\bf OpenDRIVE/OpenSCENARIO Generation Step}: The AutoPASCAL first reads the synthetic cycle data and generates the OpenDRIVE/OpenSCENARIO files. The tool provides multiple scenario files for one map to test with the different traffic densities and the environmental configurations.
  \item {\bf RNDF/MDF Conversion Step}: The AutoPASCAL reads the generated OpenDRIVE file and outputs the RNDF/MDF files used by the iREAD system. The two map formats, OpenDRIVE and RNDF, represent road networks in significantly different ways. The OpenDRIVE relies on a collection of the edges, each of which is expressed as an initial point with heading, length, and curvature. On the other hand, RNDF represents road networks with a collection of points in space, annotated using lane width and number to describe a road segment. To maintain mutual consistency, our tool subdivides the road segment into small tiles, which are approximately 3-meter in length.
\end{enumerate}

We use these four map/scenario files to conduct in-depth simulation, and our AutoPASCAL provides all the files needed to run the simulation.

In addition, for the terrain and buildings in the VIRES VTD traffic environment simulator, we use the Road Designer tool (ROD) that is a part of the VIRES VTD software.
The terrain and buildings in the traffic environment simulator do not affect in-depth simulation functionally, and these files are used only for the visualization.
The ROD tool reads the generated OpenDRIVE and outputs the OpenDRIVE and OSGB (OpenSceneGraph Binary) file. The updated OpenDRIVE file contains not only the road networks but also the buildings. Also, the OSGB describes the terrain around the road networks.

We present two examples of the synthetic cycles that are developed by the AutoPASCAL in Figures \ref{fig:NRELA_visualization} and \ref{fig:NRELB_visualization}.
Figure \ref{fig:NRELA_visualization} presents a short synthetic scenario with a total length of 1.6 (km) and 5 road intersections, of which 2 are controlled by traffic lights and 3 are controlled by stop signs.
Figure \ref{fig:NRELB_visualization} shows a longer synthetic scenario with a total length of 15.6 (km) and 14 road intersections, of which 8 are controlled by traffic lights and 6 are controlled by stop signs.
Here, Figures \ref{fig:NRELA_visualization}-(a) and \ref{fig:NRELB_visualization}-(a) are visualized from the OpenDRIVE in the VIRES traffic environment simulator.
Figures \ref{fig:NRELA_visualization}-(b) and \ref{fig:NRELB_visualization}-(b) are generated from the RNDF in the iREAD system.
By using the AutoPASCAL, the VIRES simulator and the iREAD system maintain the consistency of the geographical features, such as the location of the intersections, road curvature, lane width, lane number, and road grade.

\begin{figure}[!t]
\centering
\includegraphics[width=8.3cm]{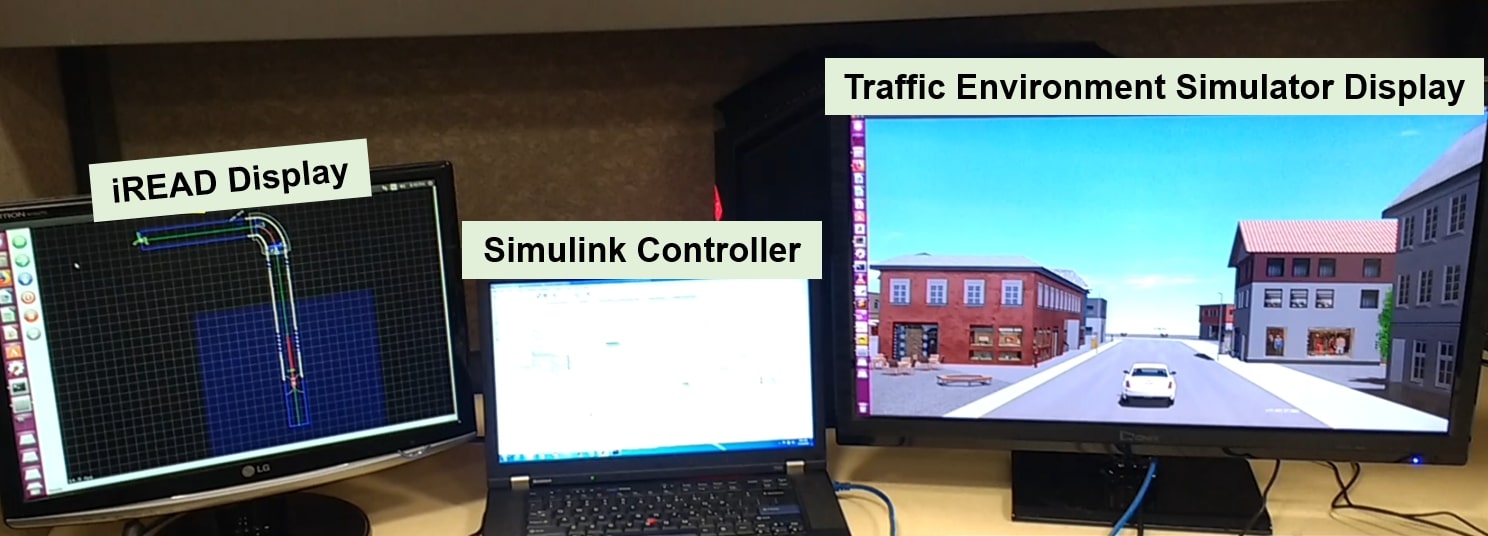}
\caption{InfoRich Co-simulation Hardware Setup.}
\label{fig:hardwaresetup}
\end{figure}

\begin{figure}[!b]
\centering
\includegraphics[width=4.45cm]{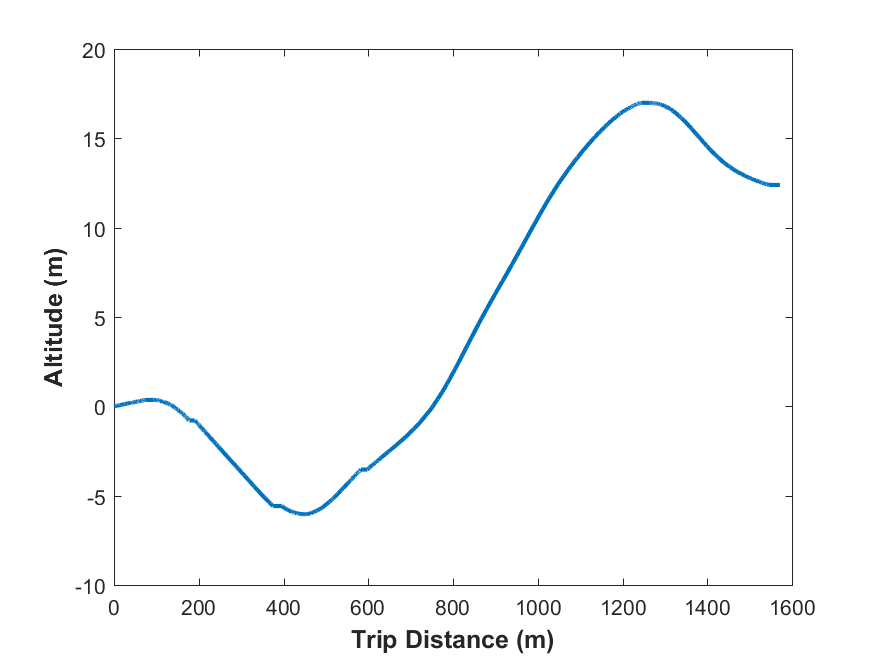}
\caption{Altitude Changes in the Synthetic Cycle.}
\label{fig:altitudechanges}
\end{figure}

 \vspace{0.1in}

\section{INFORICH: PLATFORM AND ECO-DRIVING STRATEGIES}

In this section, we review the generated synthetic scenarios and the data generated from the co-simulation platform. In addition, we introduce the {\it InfoRich} driver concept that improves the energy efficiency.

\subsection{Generated Synthetic Scenarios}


We utilize 20 synthetic cycles, including 10 short cycles and 10 long cycles derived from TSDC driving cycle data.
Table \ref{table:NRELB_visualization} presents the statistics for the cycle distance, the number of Traffic Lights (TLs), and the number of Road Intersections (RIs).
In the data set, the longest cycle has 20.3 (km) and the shortest cycle has 1.53 (km).
The longest cycle has 12 traffic lights, 21 turns, and overall 31 road intersections.



\begin{figure}[!b]
\centering
\includegraphics[width=5.25cm]{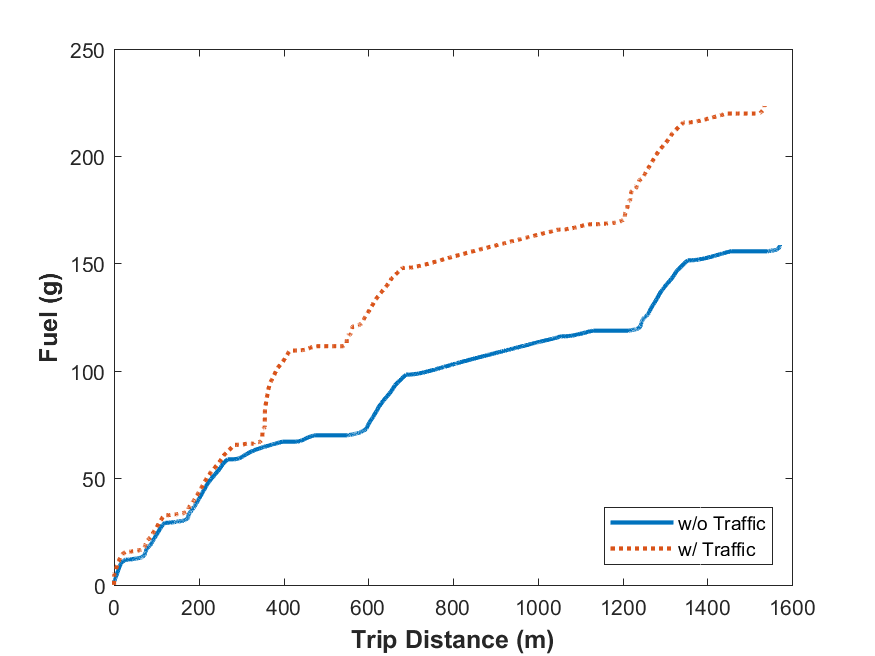}
\caption{Total Fuel Consumption in the Synthetic Cycle.}
\label{fig:totalfuelconsumption}
\end{figure}

\begin{figure*}[!t]
  \begin{center}
    \begin{tabular}{c}
      \begin{minipage}{0.33\hsize}
        \begin{center}
          \includegraphics[width=5.35cm]{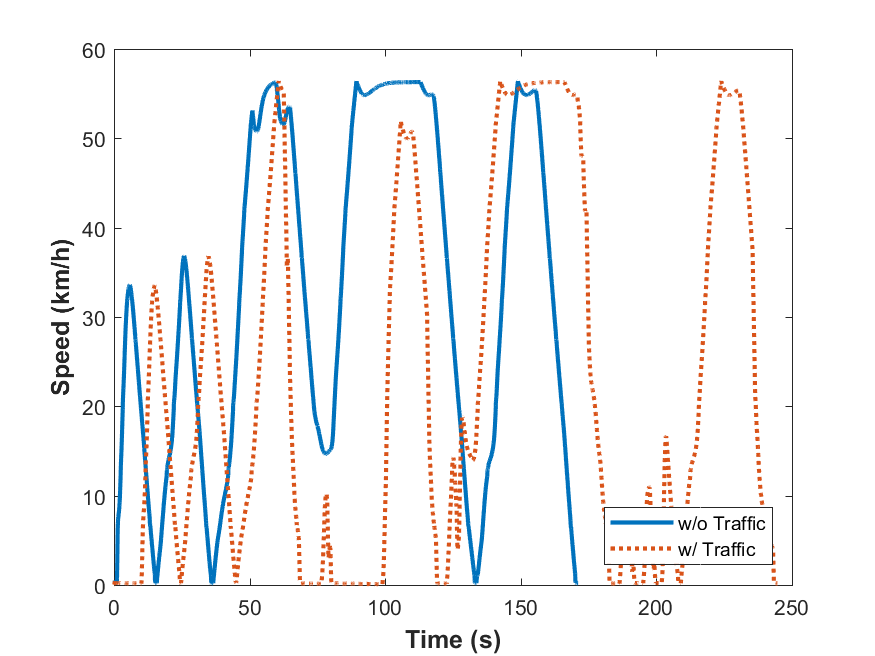}
	\hspace{6.8cm} (a) Speed Changes.
        \label{fig:distance_forsim}
        \end{center}
      \end{minipage}
      \begin{minipage}{0.33\hsize}
        \begin{center}
          \includegraphics[width=5.35cm]{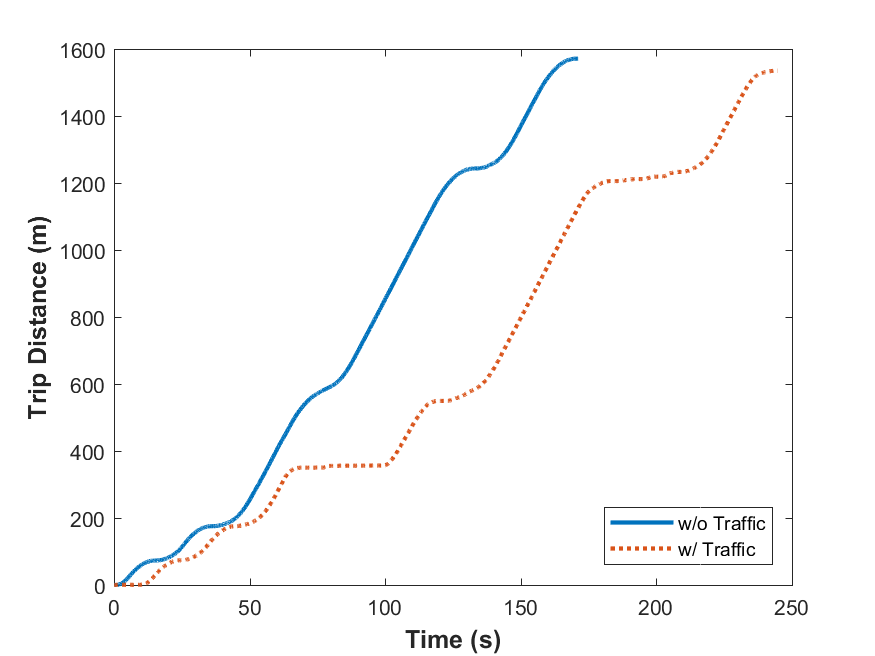}
	\hspace{6.8cm} (b) Travel Distance.
          \label{fig:fuelRate}
        \end{center}
      \end{minipage}
      \begin{minipage}{0.33\hsize}
        \begin{center}
          \includegraphics[width=5.35cm]{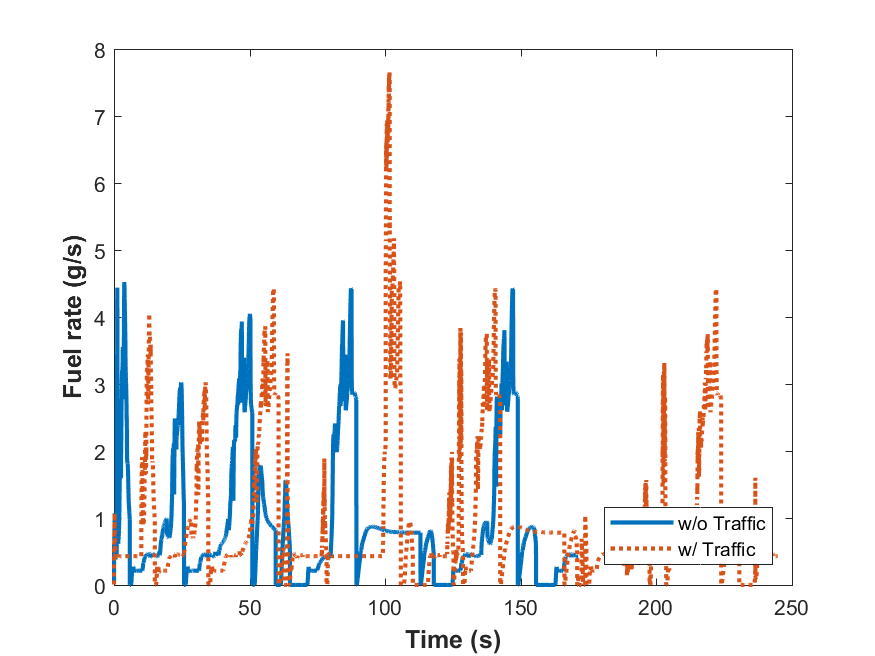}
	\hspace{6.8cm} (c) Fuel Rates.
          \label{fig:fuelRate}
        \end{center}
      \end{minipage}
    \end{tabular}
    \caption{Recorded Data of the Ego Vehicle in the Synthetic Cycle.}
\label{fig:evaluation_distance_fuel}
  \end{center}
\end{figure*}

\vspace{0.1in}

\subsection{Profile for Baseline Driver}

To evaluate the fuel consumption for a baseline driver in the InfoRich co-simulation platform, as shown in Figure \ref{fig:hardwaresetup}, we utilize the synthetic cycle that is shown in Figure \ref{fig:NRELA_visualization} discussed in Section \ref{Map_and_Scenario}.
We evaluate the fuel rate, vehicle speed, and total consumed fuel for two scenarios: (i) Without Traffic and (ii) With Traffic.
The traffic light cycles are arbitrarily fixed and synchronized: Red for 15 seconds, Green for 13 seconds, and Yellow for 2 seconds.

First, in Figure \ref{fig:altitudechanges}, we present the observed altitude changes of the travel, where the values represent the relative altitude against the Origin of the trip.
In the synthetic cycle, the Ego vehicle first goes down and then climbs an uphill until its altitude reaches 17 (m).
The road grade is a significant factor to determine the fuel consumption and to test our eco-driving applications.
These observed data is very similar to the synthetic cycle, and it shows the capability that the co-simulation platform provides the realistic road networks.



\begin{figure}[!t]
\centering
\includegraphics[width=5.80cm]{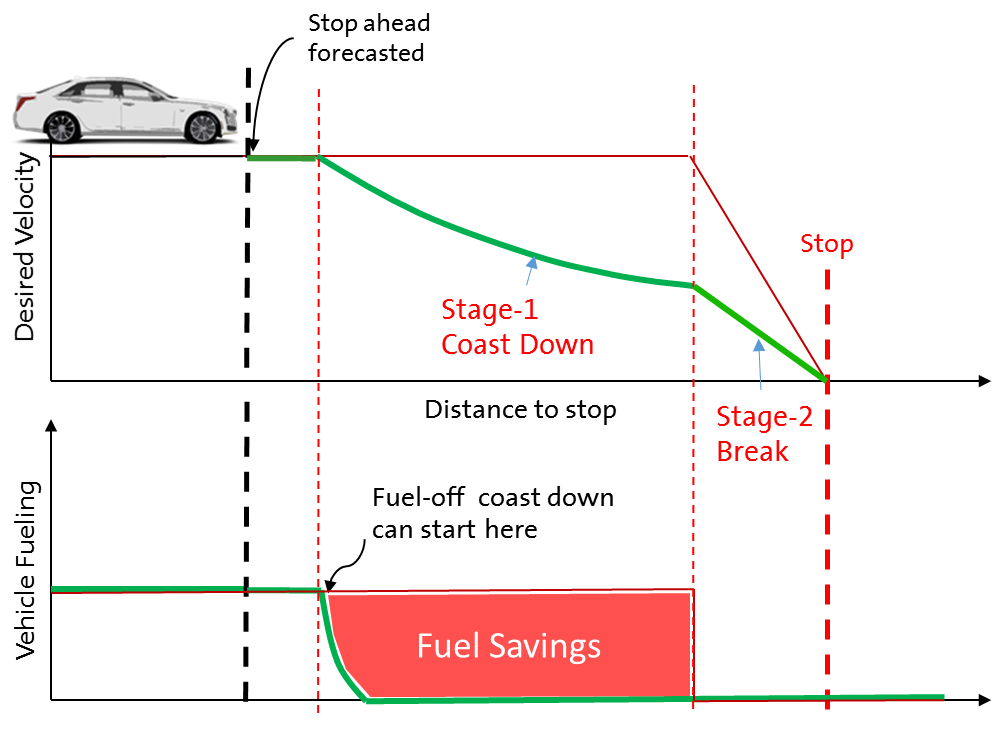}
\caption{{\it InfoRich} Eco-Approach Strategy.}
\label{fig:EcoApproach}
\end{figure}

Next, we present the total fuel consumption as shown in Figure \ref{fig:totalfuelconsumption}, and we also show the speed changes, traveled distance, and fuel rate in Figures \ref{fig:evaluation_distance_fuel}-(a), \ref{fig:evaluation_distance_fuel}-(b), and \ref{fig:evaluation_distance_fuel}-(c), respectively.
When there is traffic, the Ego vehicle will slow down to maintain an appropriate headway.
The Ego vehicle takes longer trip time and more opportunities to accelerate when it has traffic.
Due to the traffic around the Ego vehicle, the fuel consumed in the same route has been increased more than 40 \% in this synthetic cycle.
With the InfoRich Eco-Autonomous Driving (iREAD) System including multiple eco-driving applications, we can minimize the impact of the traffic on the fuel consumption.

In this evaluation, we show the capabilities of the simulation environment and study the impact of the traffic on the fuel efficiency.
The InfoRich co-simulation platform has more factors we can control, such as the pedestrians, the SPaT (Signal Phase and Timing), and the sensor models.
For example, the reliable sensor coverage distances may decrease due to the bad weather, such as fog or heavy snow.
By changing the sensor models, the InfoRich co-simulation platform enables us to study the fuel efficiency and safety of the eco-driving applications under these severe conditions.

In addition, the baseline driver simply uses the V2I communications to know the current state of the traffic light.
By using the SPaT (Signal Phase and Timing) messages of the V2I communications \cite{kenney2011dedicated}, the vehicle can access the detailed information of the SPaT defined by the SAE J2735 standard \cite{J2735}, such as the remaining time for the current state, the time duration for the next state, and the time duration for the entire cycle.
These information might help to decrease the fuel consumption.

%
%
%

\vspace{0.1in}

\subsection{Eco-Driving Strategies}

The InfoRich strategies include three eco-driving applications: Eco-Approach, Eco-Departure, and Eco-Cruise.

The Eco-Approach is defined as the energy-efficient vehicle and powertrain operation that aggressively applies coasting to bring a vehicle to a stop. As shown in Figure \ref{fig:EcoApproach}, the vehicle save the energy by predicting the stop in advance and by proactively engaging in coasting to efficiently convert the vehicle kinetic energy to the distance traveled.

\begin{figure}[!t]
\centering
\includegraphics[width=5.80cm]{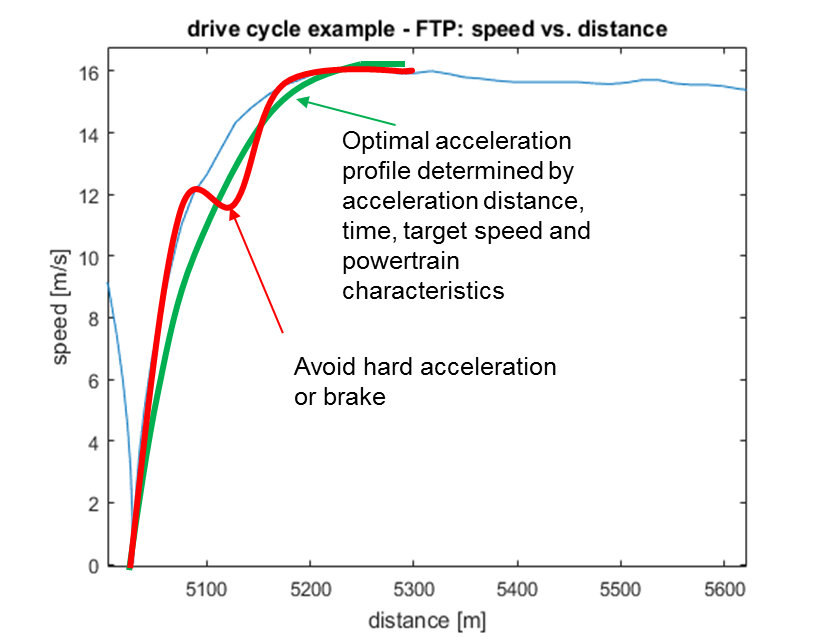}
\caption{{\it InfoRich} Eco-Departure Strategy.}
\label{fig:EcoDeparture}
\end{figure}

The Eco-Departure is the energy-efficient vehicle maneuver for a smooth and less-aggressive acceleration, as shown in Figure \ref{fig:EcoDeparture}.
The target cruise speed and the distance to reach the speed typically depend on the traffic preview information that are available from on-board sensors and vehicular communications.
Meanwhile, off-line optimization is required to obtain the most energy-efficient acceleration profile that reach the target speed and distance.

\begin{figure}[!t]
\centering
\includegraphics[width=5.80cm]{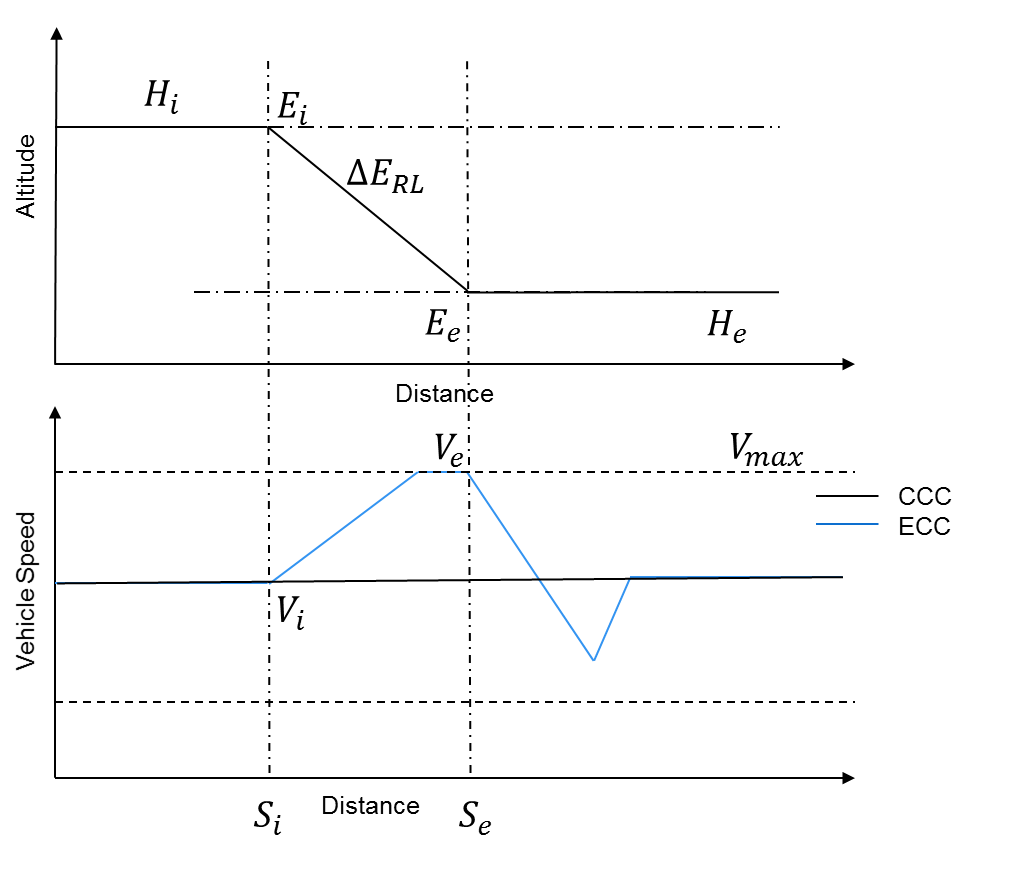}
\caption{{\it InfoRich} Eco-Cruise Strategy.}
\label{fig:EcoCruise}
\end{figure}

In addition, the Eco-Cruise is the vehicle velocity control strategy to save the fuel consumption.
One prospective application is an Eco-Cruise Control (ECC), as shown in Figure \ref{fig:EcoCruise}, which accounts for the road grade and curvature based on the Conventional Cruise Control (CCC).
Unlike the CCC, the ECC determines a flexible velocity band based on the reference vehicle velocity, e.g., given by the road speed limit.
By using the information from the on-board sensors, the Eco-Cruise control module computes an optimal velocity profile within the velocity band of the reference speed, which is then interpreted as actuation commands for throttle, brake, and gear.

Overall, the InfoRich co-simulation platform with the real-world synthetic scenarios accommodates the design, test, and development of the eco-driving strategies for the realistic vehicle models.

\section{SUMMARY AND FUTURE WORK}


In this paper, we designed and developed a co-simulation platform for testing and verifying eco-driving Connected and Automated Vehicles (CAVs).
We described three main components of the co-simulation environment, and detailed each component: the InfoRich Eco-Autonomous Driving (iREAD) System, a vehicle dynamics/powertrain model, and a traffic environment simulator.
We also presented synthetic drive cycles and developed a tool for developing map and scenario files for the co-simulation environment.
We conclude that the InfoRich co-simulation platform provides a realistic simulation environment, including a vehicle dynamics model, a powertrain model, a traffic model, and a road-network model.
In future work, we will study and test the control and planning strategies for eco-driving CAVs with the co-simulation platform.
In addition, we will extend the co-simulation platform for developing a Hardware-in-the-Loop (HIL) simulator by integrating with a real vehicle.

\section*{Acknowledgment}
This paper is based upon the work supported by the United States Department of Energy (DOE), ARPA-E NEXTCAR program under award No. DE-AR0000797.
The authors would like to thank Jeff Gonder and Michael O'Keefe at National Renewable Energy Laboratory (NREL) for providing real world driving data and analysis.

\bibliographystyle{ieeetr}
\bibliography{IEEEabrv,bib_From20141215}

\end{document}